\documentclass[final,5p,twocolumn]{elsarticle}

\usepackage[tight]{subfigure}
\usepackage{array}
\usepackage{mathtools}
\usepackage{graphicx}
\usepackage{yfonts}
\usepackage{enumitem}
\usepackage{tabularx}
\usepackage{tensor}
\usepackage{multirow}
\usepackage{color}
\usepackage{amsmath,amssymb,amsthm}
\usepackage{thmtools}
	\declaretheoremstyle[
		spaceabove=\topsep,
		spacebelow=\topsep,
		headfont=\normalfont\bfseries,
		notefont=\normalfont\bfseries,
		notebraces={(}{)},
		bodyfont=\normalfont,
		postheadspace=1em,
		qed=$\square$
	]{definition}
	\declaretheoremstyle[
		spaceabove=\topsep,
		spacebelow=\topsep,
		headfont=\normalfont\bfseries,
		notefont=\normalfont\bfseries,
		notebraces={(}{)},
		bodyfont=\normalfont,
		postheadspace=1em,
		qed=$\blacksquare$
	]{proposition}
	\declaretheoremstyle[
		spaceabove=\topsep,
		spacebelow=\topsep,
		headfont=\normalfont\itshape,
		notefont=\normalfont,
		notebraces={(}{)},
		bodyfont=\normalfont,
		postheadspace=1em,
		qed=$\bigcirc$
	]{remark}
	\declaretheorem[style=definition]{definition}
	\declaretheorem[style=proposition]{proposition}
	\declaretheorem[style=remark]{remark}
\usepackage{bm} 
\usepackage{lipsum}
\usepackage{hyperref}
\hypersetup{unicode=false,          
colorlinks=true,       
linkcolor=hyperrefLinkColor,          
citecolor=hyperrefCiteColor,        
filecolor=blue,      
urlcolor=hyperrefURLColor}           
\definecolor{hyperrefLinkColor}{rgb}{0.35,0.0,0}
\definecolor{hyperrefCiteColor}{rgb}{0,0.35,0}
\definecolor{hyperrefURLColor}{rgb}{0,0,0.35}

\usepackage{mycommands}

\journal{Journal of Magnetism and Magnetic Materials} 

\begin{document}

\begin{frontmatter}



\title{Modeling spin magnetization transport\\ in a spatially varying magnetic field}


\author[address:me]{Rico A.\,R. Picone}
\address[address:me]{Department of Mechanical Engineering, University of Washington, Seattle, USA}

\author[address:me]{Joseph L. Garbini}

\author[address:or]{John A. Sidles}
\address[address:or]{Department of Orthop\ae{}dics, University of Washington, Seattle, USA}


\begin{abstract}
We present a framework for modeling the transport of any number of globally conserved quantities in any spatial configuration and apply it to obtain a model of magnetization transport for spin-systems that is valid in new regimes (including high-polarization). The framework allows an entropy function to define a model that explicitly respects the laws of thermodynamics.  Three facets of the model are explored.  First, it is expressed as nonlinear partial differential equations that are valid for the new regime of high dipole-energy and polarization.  Second, the nonlinear model is explored in the limit of low dipole-energy (semi-linear), from which is derived a physical parameter characterizing separative magnetization transport (SMT).  It is shown that the necessary and sufficient condition for SMT to occur is that the parameter is spatially inhomogeneous.  Third, the high spin-temperature (linear) limit is shown to be equivalent to the model of nuclear spin transport of Genack and Redfield~\cite{Genack1975}.  Differences among the three forms of the model are illustrated by numerical solution with parameters corresponding to a magnetic resonance force microscopy (MRFM) experiment~\cite{Degen2009,Kuehn2008,Sidles2003,Dougherty2000}.  A family of analytic, steady-state solutions to the nonlinear equation is derived and shown to be the spin-temperature analog of the Langevin paramagnetic equation and Curie's law.  Finally, we analyze the separative quality of magnetization transport, and a steady-state solution for the magnetization is shown to be compatible with Fenske's separative mass transport equation~\cite{Fenske1932}.  
\end{abstract}


\end{frontmatter}


\section{Introduction}
Magnetization transport for a spin-system in a spatially varying magnetic field has been studied theoretically~\cite{Genack1975} and experimentally~\cite{Eberhardt2007,Budakian2004}.  High-temperature models of spin magnetization transport, such as that of Genack and Redfield, do not apply for systems with high-polarization.  With recent significant enhancements of the technique of \emph{dynamic nuclear polarization} (DNP)~\cite{Ni2013,Abragam1978,Krummenacker2012}, which has been shown to achieve significant \emph{hyperpolarization}, models that can describe the high-polarization regime in a spatially varying magnetic field are needed.



We present a framework for modeling the transport of any number of globally conserved quantities in any spatial configuration. We then apply it to obtain a model of magnetization transport for spin-systems that is valid in new regimes (including high-polarization). Finally, we analyze the separative quality of the magnetization transport. A particularly useful feature of the framework is that specifying an entropy density function \emph{completely determines the system model}. Such a function is presented for a spin-system, and its validity is demonstrated by deriving classical models from it, which is justification for using it in new regimes (as we do in \autoref{sec:magnetizationtransport}).  

In \autoref{sec:entropictheory}, we introduce the framework.  It is general in the sense that it can be applied to systems with any number of globally conserved distributed quantities that evolve over any (smooth) spatial geometry in any number of spatial dimensions.  The laws of thermodynamics are included \textit{a priori} such that any specific model based on the framework will be guaranteed to respect all four laws.

In \autoref{sec:magnetizationtransport}, we apply the framework by specifying conserved quantities, an entropy function, and a spatial geometry for a spin-system and thereby obtain a new model of magnetization transport in a magnetic field gradient.  It accommodates previously unmodeled regimes of high energy and high polarization, such as may develop with DNP.  The remainder of the section explores the model in various limits and connects them to previous models.

In \autoref{sec:fenske}, we analyze the separative quality of magnetization transport, highlighting the parallelism between it and the separative mass transport work that began with Fenske~\cite{Fenske1932}.  Magnetization transport in a magnetic field gradient is both diffusive and separative, and the latter is of particular interest for technologies that may be enhanced by hyperpolarization, such as magnetic resonance imaging (MRI), nuclear magnetic resonance (NMR) spectroscopy, and magnetic resonance force microscopy (MRFM).  DNP has been used to achieve significant hyperpolarization through transferring polarization from electron-spins to nuclear-spins.  But DNP is still a slow process, taking tens to thousands of seconds to develop, and it is impeded by magnetic field gradients, which for certain applications (such as MRFM) is undesirable.  We consider the feasibility of a different technique in which no polarization is transferred among spin-species, but in which magnetization is concentrated by the phenomenon of \emph{separative magnetization transport} (SMT).  We develop the necessary and sufficient conditions for the SMT of a single spin-species\,---\,most notably, that the magnetic field must be spatially varying.  Taken as a whole, this paper lays the groundwork for an investigation into how the SMT-effect might be enhanced to produce hyperpolarization.

\section{Framework for transport analysis} \label{sec:entropictheory}
We will proceed in the coordinate-free language of differential geometry, which allows the laws of thermodynamics to be respected explicitly, regardless of spatial geometry or the number of conserved quantities.

What follows is a necessarily extensive list of definitions and remarks. As we will see, the mathematical rigor of these definitions will enable and greatly simplify the subsequent theoretical development.

%
%
%

The following definitions are the elements from which two propositions are constructed that describe a framework for transport analysis and its adherence to the laws of thermodynamics. In any specific application of the framework, defining the spatial geometry, conserved quantities, entropy function, and a space-time scale will be sufficient to construct a model of transport that respects the laws of thermodynamics from the following definitions (as detailed in Remark~\ref{rem:remainingspecs}).

The elements of the framework are defined in the following order:
\begin{enumerate}[parsep=0pt,label=(\alph*)]
	\item the spatial manifold, metric, and coordinates; 
	\item conserved quantities and their local densities;
	\item the entropy density and thermodynamic potentials;
	\item Onsager's kinetic coefficients;
	\item the current of local quantity densities;
	\item the continuity equation for local densities; and
	\item a transport rate tensor and an \textit{ansatz} further specifying the kinetic coefficients.
\end{enumerate}
\autoref{prop:laws} will describe how the laws of thermodynamics are satisfied in the definitions and \autoref{prop:framework} will assert that the definitions describe a physically valid model of transport. We begin with spatial considerations.

\begin{definition}[spatial manifold]\label{def:spatialmanifold}
\noindent Let $\U$ be a Riemannian (smooth) manifold, which represents the spatial geometry of a macroscopic thermodynamic system. We call $\U$ the \emph{spatial manifold}.
\end{definition}
%
\noindent For many applications, a Euclidean space\footnote{See~\cite{Bullo2005}, p.~22 and~\cite{Lee2012}, p.~598.} $\mathbb{R}^m$ is an appropriate choice for $\U$.

\begin{figure}[tb]
	\centering
	\includegraphics[width=0.6\linewidth]{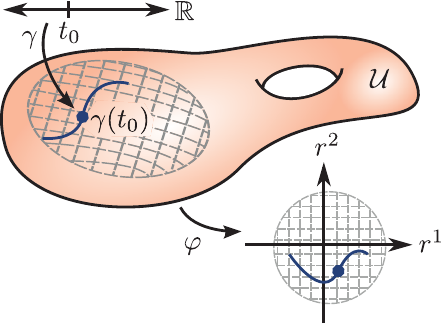}
	\caption{Local spatial coordinates $(r^\alpha)$ and coordinate map $\varphi$ to $\mathbb{R}^m$ (shown here with $m = 2$). A curve $\gamma(t)$ is a map from an interval on $\mathbb{R}$ to the spatial manifold $\U$ and has coordinate representation $\varphi(\gamma(t))$.}
	\label{fig:coordinates}
\end{figure}


\begin{remark}[spatial coordinates]\label{rem:spatialcoordinates}
	Let $\varphi:\U\rightarrow\mathbb{R}^m$ be some \emph{local coordinate map}.\footnote{See~\cite{Lee2012}, pp.~15-16, 60-65.} Typically, we will denote component functions of $\varphi$, defined by $\varphi(p) = \left(r^1(p),\ldots,r^m(p)\right)$ for some point $p\in\U$, as $(r^1,\ldots,r^m)$. These are called \emph{local spatial coordinates}, and typically denoted $(r^\alpha)$. (See \autoref{fig:coordinates}.)
\end{remark}


By definition, the Riemannian spatial manifold $\U$ is endowed with a Riemannian metric, which determines the geometry of $\U$.

\begin{definition}[spatial metric]\label{def:spatialmetric}
\noindent Let $g$ be a \emph{Riemannian metric}\footnote{See~\cite{Lee1997}, p.~23.} on $\U$. We call $g$ the \emph{spatial metric}.
\end{definition}

\noindent In local spatial coordinates, the metric is written as
\begin{align}\label{eq:metric}
	g = g_{\alpha\beta}\ dr^\alpha \otimes dr^\beta.
\end{align}

\begin{definition}[conserved quantities]\label{def:conservedquantities}
\noindent Let $\qvec\in\mathbb{R}^n$ be the $n$-tuple $\qvec = (q_1,\ldots,q_n)$, where $q_i\in\mathbb{R}$ represents a \emph{conserved quantity}.
\end{definition}

\begin{definition}[standard thermodynamic dual basis]\label{def:standardcovectorbasis}
\noindent Let the ordered basis $(\varepsilon^1,\ldots,\varepsilon^{n})$ for $\mathbb{R}^n$ be
	\begin{align*}
		\varepsilon^1 = (1,0,\ldots,0), & &
		\ldots & &
		\varepsilon^{n} = (0,0,\ldots,1).
	\end{align*}
	We call this the \emph{standard thermodynamic dual basis},\footnote{Although it is an uncommon practice to introduce a dual basis before a basis, we do so here because the quantities represented by the quantities $\qvec$ and $\rhovec$ are more naturally\,---\,from a physical standpoint\,---\,considered dual to the potentials $\Omegavec$. Yet, the quantities naturally arise first in the series of definitions.} and it is often denoted $(\varepsilon^i)$.
\end{definition}
\noindent In the standard basis, with the \emph{Einstein summation convention},
\begin{align}
	\qvec = [q_\varepsilon]_i \varepsilon^i.
\end{align}

\begin{definition}[local quantity density]\label{def:quantitydensities}
\noindent Let $\mathcal{O}^*$ be the set of smooth maps from $\U\times\mathbb{R}$ (where $\mathbb{R}$ represents time) to $V^*\equiv \mathbb{R}^n$ (i.e. for each point in space and time we assign a vector in $\mathbb{R}^n$). Given a vector of conserved quantities $\qvec$, let $\rhovec\in\mathcal{O}^*$ represent the local spatial density of each of the conserved quantities $\qvec$, such that
	\begin{align}\label{eq:qrho}
		\qvec = \int_\U \rhovec\ dv,
	\end{align}
where $dv$ is a volume element of $\U$.
\end{definition}

\begin{figure}[tb]
	\centering
	\includegraphics[width=0.6\linewidth]{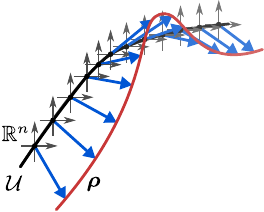}
	\caption{$\rhovec$ considered as a time-varying section of a product bundle with (black) base $\U$ and (red) section $\rhovec$. The (gray) vector space $V^*=\mathbb{R}^n$ is duplicated at each point on $\U$. Each copy of $\mathbb{R}^n$ corresponds to a (blue) arrow that represents $\rhovec$ at that point in space and at some given time. In this example, $\U$ is one-dimensional (with curvature) and the vector space is $\mathbb{R}^2$, meaning there is a single spatial dimension and there are two conserved quantities.}
	\label{fig:rhobundle}
\end{figure}

\noindent In the standard thermodynamic dual basis $(\varepsilon^i)$, we write
\begin{align}
	\rhovec = [\rho_\varepsilon]_i \varepsilon^i,
\end{align}
where each $[\rho_\varepsilon]_i$ is a function $[\rho_\varepsilon]_i:\U\times\mathbb{R}\rightarrow \mathbb{R}$. The mathematical structure of $\rhovec$ assigned in the definition is equivalent to a section of the product bundle $\U\times\mathbb{R}\times V^* \rightarrow \U\times\mathbb{R}$. \autoref{fig:rhobundle} illustrates this description, where copies of $V^*=\mathbb{R}^n$ correspond to each location in $\U$. 

We now turn to entropic considerations.



\begin{definition}[local entropy density]\label{def:entropy}
\noindent Let the \emph{local entropy volumetric density function} \mbox{$s:V^*\rightarrow\mathbb{R}$} be a function that is nonnegative and concave.
\end{definition}

	The restriction of the local entropy density $s$ to nonnegative functions satisfies the \emph{third law of thermodynamics}. Moreover, we require that $s$ be concave to allow the Legendre dual relationship that will now be introduced.

At times it is convenient to work with another set of variables called \emph{local thermodynamic potentials}. These are significant because their spatial gradients drive the flow of $\rhovec$.

\begin{definition}[local thermodynamic potentials]\label{def:thermodynamicpotentials}
\noindent Let $\Omegavec:\U\times\mathbb{R}\rightarrow V$ be defined by the relation
	\begin{align}\label{eq:mu}
		\Omegavec = ds \circ \rhovec,
	\end{align}
	where the exterior derivative $d$ is taken with respect to the vector space $V^*$.\footnote{The vector space $V^*$ has dual space $V^{**} = V$. Therefore $\Omegavec \in \mathcal{O}$ maps to vectors in $V$ and $\rhovec \in \mathcal{O}^*$ maps to covectors in $V^*$.}
	We call $\Omegavec$ the \emph{local thermodynamic potential}, and it is the Legendre transform\footnote{For an excellent article on the Legendre transform and this duality, see~\cite{Zia2009}.} $n$-tuple conjugate of $\rhovec$. The dual space of $\mathcal{O}^*$ is denoted $\mathcal{O}$, and so $\Omegavec\in\mathcal{O}$. The duality gives the \emph{standard thermodynamic basis} $(E_i)$ to be such that $E_i(\varepsilon^j) = \delta_i^j$, where $\delta$ is the Kronecker-delta.
\end{definition}

\begin{figure}[tb]
	\centering
	\includegraphics{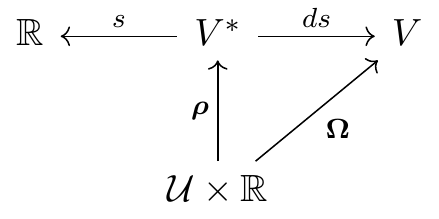}
	\caption{A commutative diagram relating the spacetime manifold $\U\times\mathbb{R}$, to thermodynamic quantity densities $\rhovec\in\mathcal{O}^*$, thermodynamic potentials $\Omegavec\in\mathcal{O}$, and the entropy density function $s\in C^\infty(V^*,\mathbb{R})$.}
	\label{fig:commutativeDiagram1}
\end{figure}

The standard thermodynamic basis representation of the potential is
\begin{align}
	\Omegavec = [\Omega_E]^i E_i.
\end{align}
The convention has been adopted that thermodynamic vectors are represented by uppercase symbols with upper indices on vector components (e.g. $[\Omega_E]^i$) and thermodynamic covectors are represented by lowercase symbols with lower indices on covector components (e.g. $[\rho_\varepsilon]_i$).


Commonly, inverse temperatures are the thermodynamic potentials of internal energy quantities. In this manner, other thermodynamic potentials can be considered to be analogs of inverse temperature. For instance, magnetic moment quantities have spin-temperature thermodynamic potentials. Keeping this in mind can aid the intuition that spatial gradients in $\Omegavec$ drive the flow of $\rhovec$, as in the familiar case of heat transfer being driven by gradients in (inverse) temperature.



We now begin to construct the current of conserved quantity densities $\jvec$. First, a discussion of spatially and thermodynamically indexed tensor structures is needed, and this requires calculus on manifolds.\footnote{See both~\cite{Lee2012} and~\cite{Spivak1965}.}

\begin{figure}[tb]
	\centering
	\includegraphics[width=.6\linewidth]{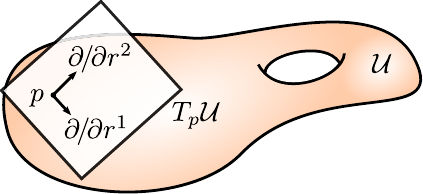}
	\caption{The tangent space $\TpU$ at point $p\in\U$ with the standard basis $\partial/\partial r^\alpha|_p$.}
	\label{fig:tangentspace}
\end{figure}

In order to do calculus on manifolds, the notion of a \emph{tangent space} is required:\footnote{See~\cite{Lee2012}, p.~54.} a tangent space at a point $p$ on the spatial manifold is a vector space $\TpU$ on which tangent vectors of curves through $p$ live (see \autoref{fig:tangentspace}). A chart that includes $p$ provides a convenient basis for $\TpU$ via its \emph{coordinate vectors}\footnote{See~\cite{Lee2012}, p.~60.} at point $p$, $\partial/\partial r^\alpha|_p$, where $(r^\alpha)$ is the local coordinate representation of $p \in \U$. The \emph{tangent bundle} $\TU$ is the disjoint union of the tangent spaces at all points on the manifold.\footnote{See~\cite{Lee2012}, pp.~65-8.} 

The \emph{dual space of the tangent space at $p$}, or \emph{cotangent space} $\TpStarU$, can be given a convenient basis $dr^\alpha|_p$ by dual-mapping the tangent space basis $\partial/\partial r^\alpha|_p$ to the cotangent space. The \emph{cotangent bundle} $\TStarU$ is the disjoint union of the cotangent spaces at all points on the manifold.\footnote{See~\cite{Lee2012}, pp.~272-303}.

Spatially indexed structures will be expressed in terms of the coordinate vectors $\partial/\partial r^\alpha|_p$ and coordinate covectors $dr^\alpha|_p$. Tensors that are indexed by both spatial coordinates and thermodynamic bases we call \emph{thermometric structures}.

The following convention for describing thermometric structures will be used. As with any tensor or tensor field, the order of their indices is merely conventional, but must be accounted. No convention is established for the ordering of the indices, but we will describe a tensor as being indexed covariantly (acting on covectors) or contravariantly (acting on vectors), first with a thermodynamic pair, say $(0, 2)$, and second with a spatial pair, say $(1, 1)$. For instance, a tensor at some point $p\in\U$ might be $(0, 2)\otimes(1, 1)$, which would have as its standard basis some permutation of the tensor product $dr^\alpha \otimes dr^\beta \otimes E_i \otimes \varepsilon^j$. 

The convention that has been adopted is that spatial coordinate vectors $\partial/\partial r^\alpha$ and covectors $dr^\beta$ are indexed by the lowercase Greek alphabet and thermodynamic basis vectors $E_i$ and covectors $\varepsilon^j$ are indexed by the lowercase Latin alphabet.

\begin{definition}[Onsager kinetic coefficients]\label{def:F}
\noindent Let $\F$ be defined as a positive-definite type $(0, 2) \otimes (0, 2)$ thermometric contravariant tensor field,\footnote{Note that a tensor field is a tensor bundle section.} which is called \emph{Onsager kinetic coefficients} tensor field, which serves as a thermodynamic and spatial metric.
\end{definition}

	To satisfy the \emph{second law of thermodynamics}, $\F$ must be positive semi-definite, so Definition \ref{def:F} satisfies the Second Law.

Typically, a contravariant tensor field is considered to assign a tensor map at each point $p\in\U$ from the tangent space $\TpU$ to a real number. However, considering $\Omegavec$ as a section of a fiber-bundle, it is a structure analogous to the tangent bundle $\TU$, and so $\F$ is indexed with both the usual cotangent local coordinate vectors $dr^\alpha$ and the thermodynamic dual basis $(\varepsilon^i)$. We often use the symmetric basis for $\F$, $\left(\varepsilon^i \otimes dr^\alpha\right)\otimes\left(\varepsilon^j \otimes dr^\beta\right)$.

\begin{definition}[current]\label{def:current}
\noindent Let $\jvec$ be defined as
	\begin{align}\label{eq:zerothlaw}
		\jvec &= \F \circ (d \Omegavec)^\sharp
	\end{align}
	where $\sharp$ is the \emph{musical isomorphism},\footnote{The musical isomorphisms are defined by the metric $g$ as maps between the tangent bundle $\TU$ and the cotangent bundle $\TStarU$ (See~\cite{Lee2012}, pp.~341-3 and~\cite{Lee1997}, pp.~27-9.). The $\sharp$ operator maps spatial 1-forms to vectors.}
	$d$ is the \emph{exterior derivative},\footnote{The exterior derivative is the coordinate-free generalization of the familiar differential of a function. See~\cite{Lee2012}, pp.~362-72.}
	and the symbol $\circ$ denotes a functional composition. We call $\jvec$ the \emph{spatial transport current}.
\end{definition}

	\autoref{eq:zerothlaw} asserts that the current is driven by the (generalized) gradient in thermodynamic potential $\Omegavec$, which is a multi-potential statement of the \emph{zeroth law of thermodynamics}.

When $\U = \mathbb{R}^m$, as is often the case, \eqref{eq:zerothlaw} can be written in terms of the coordinate-free $\text{grad}$ operator\footnote{See~\cite{Lee2012}, p.~368.} as
\begin{align}
	\jvec = \F \circ \text{grad } \Omegavec.
\end{align}
	
With the preceding definitions, we can introduce the governing equation for $\rhovec$, which is the fundamental equation of the framework of transport.

\begin{definition}[continuity]\label{def:continuity}
\noindent Let the \emph{continuity equation} be defined as
	\begin{align} \label{eq:firstlaw}
		\partial_t \rhovec = \dstar\jvec
	\end{align}
	where $\dstar$ is the \emph{Hodge codifferential} operator.\footnote{The Hodge codifferential maps $k$-forms to $(k-1)$-forms \cite[pp.~438-9]{Lee2012}. In \eqref{eq:firstlaw} it maps a spatial 1-form to a 0-form. This is similar to the divergence operator, except that it acts on a 1-form instead of a vector.}
\end{definition}

	Equation \ref{eq:firstlaw} is an expression of the global conservation of local quantities $\rhovec$. It states that the local quantities $\rhovec$ change with the (generalized) divergence of a current. For an energy quantity, this is the \emph{first law of thermodynamics}. 

\begin{proposition}[laws of thermodynamics]\label{prop:laws}
\noindent Each of the following statements is a necessary and sufficient condition for the adherence of the framework for transport analysis to the corresponding law of thermodynamics. In aggregate, then, they are necessary and sufficient conditions for adherence to all four laws of thermodynamics.
\begin{enumerate}[parsep=0pt,align=left]
	\item[\normalfont(zeroth)] The current {$\jvec$} satisfies the equation \mbox{$\jvec = \F \circ (d \Omegavec)^\sharp$}, as in \autoref{def:current}.
	\item[\normalfont(first)] The continuity equation is equivalent to the equation $\partial_t \rhovec = \dstar\jvec$, as in \autoref{def:continuity}.
	\item[\normalfont(second)] Onsager's kinetic coefficient tensor $\F$ is positive semi-definite, as in \autoref{def:F}.
	\item[\normalfont(third)] The local entropy density function $s$ is non-negative, as in \autoref{def:entropy}.\qedhere
\end{enumerate}
\end{proposition}

Notice that, while the laws of thermodynamics have narrowed, considerably, the possible forms of the transport equation, two elements remain indefinite, although their general structures have been prescribed: the local entropy density function $s$ and the tensor of Onsager's kinetic coefficients $\F$.

The specific form of the entropy function depends on the system, and so it is as yet necessarily unspecified. The final step, then, is to specify the form of the Onsager kinetic coefficients tensor field $\F$. Two forms are presented, the first (\autoref{def:Gamma}) is quite general, and applies to systems that have different transport rates. The second (\autoref{def:ozansatz}) is an \textit{ansatz} that can be used in certain applications that have a single transport rate, and is applied in \autoref{sec:magnetizationtransport} to model the magnetization transport of a system of one spin-species.

But, first, two more definitions are required.
\begin{definition}[covariance tensor field]\label{def:covariancetensor}
\noindent Let $\cov$ be defined as the negative-definite tensor field\footnote{See Equation 2-11 of~\cite{Onsager1953}.}
	\begin{align}\label{eq:cov}
		\cov = \left(\frac{\partial^2 s}{\partial \rho_i \partial \rho_j} \ E_i \otimes E_j\right)^{-1}.
	\end{align}
	The tensor field $\cov$ represents quantum mechanical observation processes which are related to the entropy density $s$ by the expression.\footnote{This is related to the Ruppeiner metric~\citep{Ruppeiner1979,Ruppeiner1995}.} We call $\cov$ the \emph{covariance tensor field}.
\end{definition}

This is a connection between the quantum mechanical and the macroscopic descriptions of transport. It can also be expressed in terms of a free-energy and local thermodynamic potentials, which are related to the entropy and local quantity densities by the Legendre transform. 

\begin{definition}[entropy Hessian]\label{def:entropyhessian}
\noindent Let $\G$ be defined as the type $(0, 2) \otimes (0, 2)$ thermometric contravariant tensor field
\begin{align}\label{eq:Ginv}
	&\G =
	-g \otimes \cov.
\end{align}
	We will call $\G$ the \emph{entropy Hessian}.
\end{definition}

\noindent In local coordinates and the standard thermodynamic basis,
	\begin{align}
	&\G = -\left(g_{\alpha\beta}\ dr^\alpha \otimes dr^\beta\right) \otimes \left(\frac{\partial^2 s}{\partial \rho_i \partial \rho_j} \ E_i \otimes E_j\right)^{-1}.
	\end{align}
	A symmetric standard basis for $\G$ is $\left(\varepsilon^i \otimes dr^\alpha\right)\otimes\left(\varepsilon^j \otimes dr^\beta\right)$. $\G$ is positive-definite because $g$ is positive-definite by its definition as a Riemannian metric and $\cov$ is by definition negative-definite.

\begin{definition}[transport rate tensor field] \label{def:Gamma}
\noindent Let $\Gammat$ be a $(1, 1) \otimes (1, 1)$ thermometric mixed tensor field called the \emph{transport rate tensor}, which is defined by the relation
	\begin{align}
		\F = \Gammat(\G).
	\end{align}
\end{definition}

The transport rate tensor field $\Gammat$ sets the space-time scales for transport. In general, there are $n^2 \times m^2$ transport space-times scales, but we often assume many fewer by symmetry and spatial isotropy.

\begin{proposition}[framework for transport analysis]\label{prop:framework}
\noindent Given a system of thermodynamic quantities, its covariance tensor field,\footnote{The covariance tensor field can be found by observation, by quantum simulation, or by an entropy density function satisfying \autoref{eq:cov}. The Legendre duality makes it equivalent to have a known, valid free energy-density function.} and its transport rate tensor\,---\,and if the system has no significant advective transport\footnote{We have not here considered the case of advective transport. For magnetization transport, this means that only solid-state magnetization samples are considered.}\,---\,the continuity equation of \autoref{def:continuity} and its dependent definitions describe the transport of the system over times for which global quantities $\qvec$ can be considered substantially conserved.\footnote{That is, times for which \autoref{def:conservedquantities} holds for $\qvec$.}
\end{proposition}

For certain systems (e.g. a system of spins of a single species), the following \textit{ansatz} simplifies the analysis.

\begin{definition}[OZ-\textit{ansatz}]\label{def:ozansatz}
\noindent Let $\Gammaoz$ be a real number called the \emph{Onsager-Ziegler transport coefficient} that specifies a single space-time scale. Let the \emph{Onsager-Ziegler ansatz} (OZ-\textit{ansatz}) be the following relation that specifies the Onsager kinetic coefficient tensor field (\autoref{def:F}):
	\begin{align} \label{eq:OZansatz}
		\FOZ = \Gammaoz\G. \rlap{$\qquad$\qedhere}
	\end{align}
\end{definition}

\noindent Equation \ref{eq:OZansatz} assumes a single space-time scale for all transport. This \textit{ansatz} is roughly accurate in many physical systems, but we do not assert that it is generally valid. In nuclear magnetization transport, this \textit{ansatz} is usually reasonable.\footnote{See~\cite{Genack1975}, p.~83.}

\begin{remark}\label{rem:remainingspecs}
	Therefore, to implement the transport model for any specific system, the only elements needed are:
	\begin{enumerate}[parsep=0pt,label=(\alph*)]
		\item the spatial manifold $\U$, metric $g$, and local coordinates $(r^\alpha)$;
		\item the globally conserved thermodynamic quantities that define local quantity densities $\rhovec$;
		\item the local entropy density function $s$; and
		\item the transport rate tensor $\Gammat$.\qedhere
	\end{enumerate}
\end{remark}

In \autoref{sec:magnetizationtransport} the framework of \autoref{prop:framework} is developed into a model for the specific case of magnetization transport through a process of specifying the elements described in \autoref{rem:remainingspecs}.

\section{Model of magnetization transport}\label{sec:magnetizationtransport}
In this section, we construct a specific model of magnetization transport from the framework of the last section in one spatial dimension, with a single spin-species, and (therefore) with two conserved quantities.  \autoref{tab:recipe} summarizes the steps for developing this magnetization transport model.  We begin with some definitions.

\renewcommand{\arraystretch}{1.5}
\renewcommand{\tabcolsep}{.22em}
\newcolumntype{C}{>{\centering\arraybackslash}X}%
\def\tabularxcolumn#1{m{#1}}
\begin{table*}[t]
	\centering
	{\sffamily%
	\begin{minipage}{1\textwidth}
	\begin{tabularx}{\textwidth}{ |@{}>{\centering}m{4.5em}@{}|X|X| }
		\multicolumn{3}{c}{%
			\bfseries Deriving a Model of Magnetization Transport in a Magnetic Field: A Summary
		} \\ \hline
	  	\multicolumn{1}{c}{Element} & \multicolumn{1}{c}{General Framework of Transport (Sec.~\ref{sec:entropictheory})} & 
		\multicolumn{1}{c}{1D Single-Species Magnetization Theory (Sec.~\ref{sec:magnetizationtransport})} \\ \hline
	  	spatial\linebreak{}manifold\linebreak{}$\U$ & 
		Let $\U$ be a Riemannian manifold (Def.~\ref{def:spatialmanifold}) with metric $g$ (Def.~\ref{def:spatialmetric}) and local coordinates $(r^\alpha)$ (Rem.~\ref{rem:spatialcoordinates}) that represents the spatial geometry.	& 
		\parbox{.43\textwidth}{\raggedright Let $\U$ be the reals (Def.~\ref{def:Sspatialmanifold}, one spatial dimension),\linebreak the spatial coordinate be $(r)$, and\linebreak the metric be $g = dr \otimes dr$ (Def.~\ref{eq:magnetizationmetric}, Euclidean metric).} \\ \hline
		conserved\linebreak quantities\linebreak $\qvec$ &
		Let the vector $\qvec\in\mathbb{R}^n$ represent conserved quantities (Def.~\ref{def:conservedquantities}).  Let the basis $(\varepsilon^i)$ be the standard thermodynamic dual basis for $\qvec$ (Def.~\ref{def:standardcovectorbasis}).  & 
		\parbox{.4\textwidth}{\raggedright Let\linebreak $[q_\varepsilon]_1$ represent the total magnetic energy and\linebreak $[q_\varepsilon]_2$ represent the magnetic moment (Def.~\ref{def:Squantities}).} \\ \hline
		quantity\linebreak densities\linebreak $\rhovec$ &
		Let the vector-valued function $\rhovec$ represent local quantity density functions that can be integrated over $\U$ to obtain $\qvec$ (Def.~\ref{def:quantitydensities}).  The function $\rhovec$ inherits the basis $(\varepsilon^i)$. & 
		\parbox[c]{.35\textwidth}{\raggedright Let\linebreak $[\rho_\varepsilon]_1$ represent the local energy density and\linebreak $[\rho_\varepsilon]_2$ represent the magnetization (Rem.~\ref{rem:Squantitydensities}).} \\ \hline
		entropy\linebreak density\linebreak $s$ & 
		Let $s$ be the nonnegative and concave local entropy density function of quantity densities, all of which can be expressed as basis transformations of $\rhovec$. &
		Let $s$ be the entropy of mixing, as described in Def.~\ref{def:Sentropydensity}.  It is expressed without explicit spatial dependence, which is convenient for the proceeding calculations. \\ \hline
		$\rhovec$-dual\linebreak potential\linebreak $\Omegavec$ &
		The vector of local thermodynamic potentials $\Omegavec$ is the Legendre dual variable of $\rhovec$ (Def.~\ref{def:thermodynamicpotentials}).  The standard thermodynamic basis for $\Omegavec$ is $(E_i)$, where $E_i(\varepsilon^j) = \delta_i^j$.  &
		Let $\Omegavec$ be the spin-temperatures $[\Omega_E]^i = \partial s/\partial [\rho_\varepsilon]_i$.  Henceforth, use the convenient thermodynamic basis $(e_i)$ and dual basis $(e^i)$ with the transformation of Def.~\ref{def:Sthermodynamicbasis}.
		\\ \hline
		kinetic\linebreak coef.\linebreak $\FOZ$ &
		Let the covariance tensor field be (Def.~\ref{def:covariancetensor}) \mbox{$\cov = \left(\partial^2 s/\partial \rho_i \partial \rho_j \ E_i \otimes E_j\right)^{-1}$}, let the entropy Hessian be (Def.~\ref{def:entropyhessian}) \mbox{$\G = -g \otimes \cov$}, and let the transport coefficient $\Gammaoz$.  Finally, define Onsager's kinetic coefficients using the OZ-\emph{ansatz} (Def.~\ref{def:ozansatz}) $\FOZ = \Gammaoz\G$.
		 &
		 Compute $\cov$, $\G$, and $\FOZ$.  The latter two are thermometric structures with four components.  
		 Determine the transport coefficient $\Gammaoz$ from spin-system properties such as spin-species and spin-density.  A spin-diffusion constant from the literature may be appropriate.
		 \\ \hline
		 \vspace{-1.2ex}transport\linebreak current\linebreak $\jvec$ &
		 Let the transport current $\jvec$ be $\FOZ$ acting on what is the gradient, in Euclidean space, of $\Omegavec$ (Def.~\ref{def:current}):
		 {\setlength\belowdisplayskip{0pt}
		 \setlength\abovedisplayskip{0pt}
		 \begin{align*}
		 	\jvec = \FOZ \circ (d \Omegavec)^\sharp. 
		 \end{align*}}
		  &
		 Compute the current from (Def.~\ref{def:current}) using the equation:
		 {\setlength\belowdisplayskip{0pt}
		 \setlength\abovedisplayskip{0pt}
		 \begin{alignat*}{3}
			 (d\Omegavec)^\sharp = 
			&\left( \partial_r [\Omega_e]^1 \right)\ &&e_1\otimes \partial/\partial r \\[-3pt]
			+
			&\left( 
				\partial_r [\Omega_e]^2 
				+ [\Omega_e]^1 B'/B_d
			\right)\ &&e_2\otimes \partial/\partial r. \nonumber
		\end{alignat*}}
		\\[-2ex] \hline
		governing\linebreak equation\linebreak of $\rhovec$ &
		Let the continuity equation, the governing equation of $\rhovec$, be (Def.~\ref{def:continuity}):
		{\setlength\belowdisplayskip{0pt}
		\setlength\abovedisplayskip{0pt}
		\begin{align*}
			\partial_t \rhovec = \dstar\jvec. 
		\end{align*}\vspace{-1.5ex}}
		&
		Write the continuity equation with the divergence of $\jvec$:
		{\setlength\belowdisplayskip{0pt}
		\setlength\abovedisplayskip{0pt}
		\begin{alignat*}{3}
			\dstar\jvec = 
			&\left( 
				-\partial_r [j_e]_1 
				+ [j_e]_2 B'/B_d 
			\right)\ &&e^1 \\[-3pt]
			+
			&\left( -\partial_r [j_e]_2 \right) &&e^2.  \nonumber
		\end{alignat*}\vspace{-1.5ex}}
		\\ \hline
	\end{tabularx}
	\end{minipage}
	}%
	\caption{\emph{Left-to-right}: an element (left) defined in the general framework (center) is applied to specify an element of the 1D one-species magnetization model (right).  \emph{Top-to-bottom}: a summary of the derivation of the magnetization transport model.}
	\label{tab:recipe}
\end{table*}

\begin{definition}[spatial manifold]\label{def:Sspatialmanifold}
	Let the manifold $\U$ be defined by \mbox{$\U = \mathbb{R}$}. We call $\U$ the spatial manifold.\footnote{We consider geometries with \emph{transverse isotropy} of magnetization, external magnetic field, and sample composition; therefore, only one spatial dimension is of consequence. For this reason, although volumes are considered three-dimensional, in all other cases we consider only the single spatial dimension.}
\end{definition}
\noindent An atlas for $\U$ is given by the chart $\varphi:\U\rightarrow\mathbb{R}$, where $\varphi$ is the identity, which yields the single Cartesian spatial coordinate $(r)$ in the direction normal to the isotropic plane.

\begin{definition}[spatial metric]\label{eq:magnetizationmetric}
	Let $g$ be the (Riemannian) \textit{Euclidean metric}, \mbox{$g = dr \otimes dr$}.
\end{definition}

For the transport of a single spin-species, there are two globally conserved functions on $\U$, so $n = 2$.

\begin{definition}[conserved quantities]\label{def:Squantities}
	  Let $\qvec\in{R}^n$ be a vector with components:  $[q_\varepsilon]_1\in\mathbb{R}$ representing the total magnetic energy (Zeeman and dipole) and $[q_\varepsilon]_2\in\mathbb{R}$ representing the total magnetic moment.\footnote{Note that any linear transformation of these quantities is also conserved.} 
\end{definition}

\begin{remark}[quantity densities]\label{rem:Squantitydensities}
	These definitions, along with the framework of \autoref{prop:framework}, posit a temporally varying local quantity density $\rhovec\in\mathcal{O}^*$, the components of which, in the standard basis, represent the energy volumetric density ($[\rho_\varepsilon]_1$) and the magnetization ($[\rho_\varepsilon]_2$).
\end{remark}

\subsection{Local entropy density}
The local entropy density \mbox{$s:V^*\rightarrow\mathbb{R}$} is defined in a coordinate-free manner that requires some discussion.  The definition is derived from the \emph{entropy of mixing}, which is an entropy function that describes the mixing of several nonreactive quantities.

\begin{definition}[local entropy density]\label{def:Sentropydensity}
	Let $\Delta$ be the temporally invariant volumetric spin density (spins per unit volume),\footnote{Since only a single spatial dimension is considered for $\U$, volumes will have physical dimensions $(\text{length})^1$, or meters in SI units.} $B$ be the external spatially varying magnetic field, $B_d$ be the average dipole magnetic field, and $\mu$ be the magnetic moment of an individual spin.  Additionally, let the vector-valued functionals $\bm{\Phi},\bm{\Psi}\in\mathcal{O}$ be
	\begin{subequations}
		\begin{align}
			\bm{\Phi} &= \frac{1}{ B_d \mu \Delta}\left(E_1 + B(r)\ E_2\right)
		\end{align}
		and
		\begin{align}
			\bm{\Psi} &= \frac{1}{\mu\Delta}\ E_2
		\end{align}
		and the real number $\overline{p}$ be
		\begin{align}
			\overline{p} = 
			\sqrt{
				\bm{\Phi}(\rhovec)^2 + 
				\bm{\Psi}(\rhovec)^2
				}.
		\end{align}
	\end{subequations}
	Then let the local entropy density $s$ be defined as
	\begin{align}\label{eq:Sentropy}
		s(\overline{p}) = 
		\frac12 \ln{4} 
		&+ \frac12 \left(\overline{p} - 1\right) \ln{\left(1 - \overline{p}\right)} \nonumber\\ 
		&- \frac12 \left(\overline{p} + 1\right) \ln{\left(1 + \overline{p}\right)}. \rlap{\qedhere}
	\end{align}
\end{definition}

It can be shown that \eqref{eq:Sentropy} satisfies the constraints of \autoref{def:entropy}.  Although this definition makes reference to the standard basis $(E_i)$, $\bm{\Phi}$ and $\bm{\Psi}$ can be written in any basis.

Due to the logarithmic definition of entropy density, the minimum of entropy is inconsequential, except in that it is necessarily nonnegative.  With another definition of globally conserved functions, say $2\qvec$, the minimum entropy would change, but the resulting transport is invariant.

\subsection{Model of magnetization transport}
We now present the model of magnetization transport.  To remain as general as possible, the OZ transport coefficient $\Gammaoz$ will not be assigned.  This remaining parameter will require selection based on the spin system.  Diffusion coefficients are often suitable choices, and for nuclear- and electron-spin systems, the literature contains theoretical and empirical values~\cite{Budakian2004,Eberhardt2007,Genack1975,Dougherty2000}.

\begin{proposition}[model of magnetization transport]\label{prop:magnetizationtransport}
	Let a magnetization system be such that the entropy of \autoref{def:Sentropydensity} is valid and let the system have a single spin-species such that \autoref{def:Squantities} holds.  Additionally, let the system be non-advective and let the OZ-\textit{ansatz} hold for the system.    And let the system meet the other criteria of the framework for transport analysis of \autoref{prop:framework}, using Definitions~\ref{def:Sspatialmanifold}\,--\,\ref{def:Sentropydensity} where appropriate.  Then the continuity equation of \autoref{def:continuity} and its dependent definitions describe the transport of the system of magnetization.
\end{proposition}

\subsection{Magnetization transport equation}
\autoref{prop:magnetizationtransport} defines the magnetization transport equation to be the continuity equation \eqref{eq:firstlaw} with the specific choices of spatial manifold $\U$, local quantity densities $\rhovec$, and entropy density function $s$ as given in Definitions~\ref{def:Sspatialmanifold}\,--\,\ref{def:Sentropydensity}.  \emph{This is the nonlinear magnetization transport equation.}  A family of steady-state solutions for this model will be presented in \autoref{sec:steadystatesolutions}.

Two additional forms of this equation will be presented in important limits.  These are solved numerically and compared to the nonlinear model.  Along the way, various connections to prior art will be discussed.

\subsubsection{Basis considerations}

The definitions of the preceding section allow any thermodynamic basis for the local quantity density $\rhovec$.  In a spatially varying magnetic field, it is convenient and common practice to use a thermodynamic basis that is itself spatially varying.  This is computationally advantageous and allows the continuity equation to be written in a simple component form.  Two classes of basis are now defined.

\begin{definition}[homogeneous basis]\label{def:homogeneousbases}
	Let a \emph{homogeneous basis} be a thermodynamic ordered vector- or covector-basis that is not spatially varying.
\end{definition}

\noindent The standard thermodynamic bases are spatially invariant, and so they are homogeneous.

\begin{definition}[inhomogeneous basis]\label{def:inhomogeneousbases}
	Let an \emph{inhomogeneous basis} be a thermodynamic ordered vector- or covector-basis that is spatially varying.
\end{definition}

\begin{definition}[polarization thermodynamic covector basis]\label{def:Sthermodynamicbasis}
	Let $(e^1,\ldots,e^n)$ be defined as an ordered thermodynamic covector basis by the basis transform from the standard dual basis:
	\begin{align}
		[\rho_e]_i= [P]\indices{_i^j} [\rho_\varepsilon]_j,
	\end{align}
	where $P$ is the thermodynamic $(1,1)$ mixed tensor with matrix representation
	\begin{align}\label{eq:P}
		P = \frac{1}{\mu \Delta(r)}
		\begin{bmatrix}
			1/B_d		& B(r)/B_d	\\
			0			& 1
		\end{bmatrix}.\rlap{\qedhere}
	\end{align}
\end{definition}

\begin{figure}[tb]
\centering
\includegraphics[width=.8\linewidth]{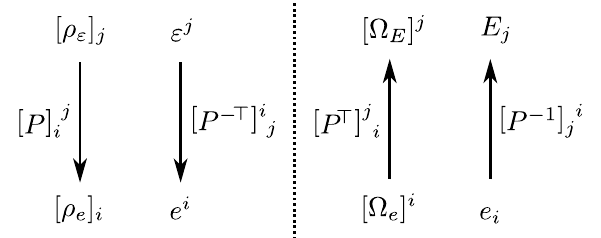}
\caption{basis transformation relations between the standard dual basis $(\varepsilon^i)$ and basis $(E_i)$ and the $e$-dual basis and $e$-basis for $\rhovec = [\rho_\varepsilon]_i \varepsilon^i = [\rho_e]_i e^i$ (left) and $\Omegavec = [\Omega_E]^i E_i = [\Omega_e]^i e_i$ (right).  The $(1,1)$ thermodynamic tensor $P$ of \autoref{def:Sthermodynamicbasis} determines the transformation.}
\label{fig:transformations}
\end{figure}

Because $\rhovec$ is a coordinate-free object, the basis vectors of $(\varepsilon^i)$ must transform to those of $(e^i)$ in a compensatory manner.  Similarly, the polarization thermodynamic vector basis for $\Omegavec$, $(e_i)$, is easily derived.  All such relations are shown in \autoref{fig:transformations}.

From the relations of \autoref{fig:transformations}, it can be shown that the basis covectors and vectors of the $e$-basis must be spatially varying, and so it is an inhomogeneous basis.

The physical interpretations of the components of $[\rho_e]_i e^i$ are no longer the same as those of $[\rho_\varepsilon]_i \varepsilon^i$.  Both components are normalized such that they are nondimensional (their basis vectors have assimilated the units originally associated with the components).  The second component has become what is typically called ``polarization,'' and ranges over the interval $[-1,+1]$.  The first component is a nondimensional version of the dipole-energy density (since the Zeeman energy density has been subtracted from the total energy density).

\subsubsection{Thermodynamic potentials and spin-temperature}

\begin{figure*}[t]
	\centering
	\subfigure[]{\includegraphics[width=.49\linewidth]{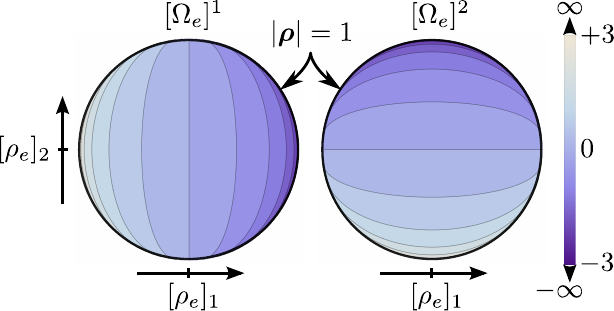}}
	\subfigure[]{\includegraphics[width=.49\linewidth]{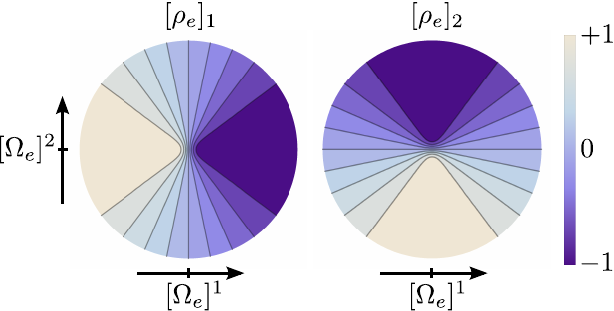}}
	\caption{(a) Density plots of $[\Omega_e]^1(\rhovec)$ (left) and $[\Omega_e]^2(\rhovec)$ (right).  As $|\rhovec|\rightarrow 1$, $[\Omega_e]^i\rightarrow\pm\infty$.  (b) Density plots of $[\rho_e]_1(\Omegavec)$ (left) and $[\rho_e]_2(\Omegavec)$ (right).}
	\label{fig:Omegarho}
\end{figure*}

Definitions \ref{def:thermodynamicpotentials} and \ref{def:Sentropydensity} give the $e$-basis representation of the thermodynamic potential
\begin{alignat}{3} \label{eq:OmegaOfRho}
	\Omegavec(\rhovec) =
	&\left( 
		-[\rho_e]_1 \text{arctanh}(|\rhovec|)/|\rhovec| 
	\right)\ &&e_1 \\
	+
	&\left( 
		-[\rho_e]_2 \text{arctanh}(|\rhovec|)/|\rhovec| 
	\right)\ &&e_2 \nonumber
\end{alignat}	
where $|\cdot|$ is the \textit{Euclidean norm}.  \autoref{fig:Omegarho} shows density plots for the components of $\Omegavec(\rhovec)$.  These approach $\pm\infty$ as $|\rhovec|$ approaches unity.

\setlength{\subfigbottomskip}{0pt}

Spin temperature is closely associated with $\Omegavec$, which can be considered to be the inverse spin-temperature associated with each quantity $\rhovec$.  In the standard basis, the components of $\Omegavec$ represent the inverse spin-temperatures of total energy-density ($[\Omega_E]^1$) and total magnetization ($[\Omega_E]^2$).  In the $e$-basis, they represent the inverse spin-temperatures of the normalized dipole energy-density ($[\Omega_e]^1$) and polarization ($[\Omega_e]^2$).

This interpretation is consistent with the relationship illustrated in \autoref{fig:Omegarho} in which as \mbox{$|\rhovec|\rightarrow 1$}, $[\Omega_e]^i\rightarrow\pm\infty$.  This means that high spin-temperatures correspond to low energy densities and polarization and low spin-temperatures correspond to high energy and polarization.

\autoref{eq:OmegaOfRho} can be solved over the domain $\tanh|\Omegavec|\leq 1$ for
\begin{alignat}{3}
	\rhovec(\Omegavec) = 
	&\left( 
		\pm [\Omega_e]^1 \tanh(|\Omegavec|)/|\Omegavec|
	\right)\ &&e^1 \label{eq:rhoomega}\\
	+
	&\left( 
		\pm [\Omega_e]^2 \tanh(|\Omegavec|)/|\Omegavec| 
	\right)\  &&e^2 \nonumber
\end{alignat}
where the terms are positive for $[\Omega_e]^1 < 0$, positive or negative for $[\Omega_e]^1 = 0$, and negative otherwise.  The density plots of \autoref{fig:Omegarho} show the functional relationships of the components. 

It is simple to show that the following two identities hold for the relationship between $\rhovec$ and $\Omegavec$ for all time and space:
\begin{subequations}
\begin{align}
	\frac{[\rho_e]_2}{[\rho_e]_1}
	&=
	\frac{[\Omega_e]^2}{[\Omega_e]^1}\ \text{and} \\
	|\rhovec| &= \tanh|\Omegavec|. \label{eq:magidentity}
\end{align}
\end{subequations}

\subsubsection{Magnetization transport equation: method}
Although there is no compact expansion of the magnetization transport equation in coordinates, the key operations required to derive it in perhaps the most compact basis, the $e$-basis, are here presented.

Let $\Delta$ be spatially homogeneous.  The first operation is the gradient of the potential $(d\Omegavec)^\sharp$ that is an element of the expression for $\jvec$ \eqref{eq:zerothlaw}.  In a single spatial dimension with coordinate $(r)$ and thermodynamic $e$-basis, this amounts to
\begin{alignat}{3}
	 (d\Omegavec)^\sharp = 
	&\left( \partial_r [\Omega_e]^1 \right)\ &&e_1\otimes \partial/\partial r \label{eq:dOmegavarepsilon} \\
	+
	&\left( 
		\partial_r [\Omega_e]^2 
		+ [\Omega_e]^1 B'/B_d
	\right)\ &&e_2\otimes \partial/\partial r \nonumber
\end{alignat}
where the field-gradient term arises from the inhomogeneity of the $e$-basis.  Second and similarly, it can be shown that the codifferential of the current is
\begin{alignat}{3}\label{eq:codifferentialj}
	\dstar\jvec = 
	&\left( 
		-\partial_r [j_e]_1 
		+ [j_e]_2 B'/B_d 
	\right)\ &&e^1  \\
	+
	&\left( -\partial_r [j_e]_2 \right) &&e^2.  \nonumber
\end{alignat}

\subsubsection{Steady-state solutions}\label{sec:steadystatesolutions}
A family of steady-state solutions is developed.  Later in the section, certain limits are explored in which familiar theories are shown to be subsets of this family of solutions.

Steady-state means that $\dstar\jvec$ is temporally invariant.  A simple solution can be developed in the case that $\jvec = 0$.  Then from \eqref{eq:zerothlaw}, it can be shown that $(d\Omegavec)^\sharp$ must be zero because $\FOZ$ is positive definite.  In the standard basis this implies that the components $\partial_r [\Omega_E]^i = 0$, i.e. the system evolves toward uniform distributions of inverse spin-temperatures.  In the polarization basis it implies that the components of \eqref{eq:dOmegavarepsilon} are zero.  Let $[\Omega_e^0]^i$ denote constants determined by a boundary condition $r_0$, and let $\overline{B}(r) = (B(r) - B(r_0))/B_d$ and $\overline{\Delta}(r) = \Delta(r)/\Delta(r_0)$. The resulting simple system of ordinary differential equations can be solved for
\begin{alignat}{3}
	\Omegavec(0,r) =
	&\left( [\Omega_e^0]^1 \overline{\Delta}(r) \right)\, 
	&&e_1 \label{eq:OmegaSS}\\
	+
	&\left( 
		[\Omega_e^0]^2
		-
		\overline{B}(r)[\Omega_e^0]^1
	\right)
	\overline{\Delta}(r)\, &&e_2. \nonumber
\end{alignat}

\begin{figure}[tb]
	\centering
	\includegraphics[width=1\linewidth]{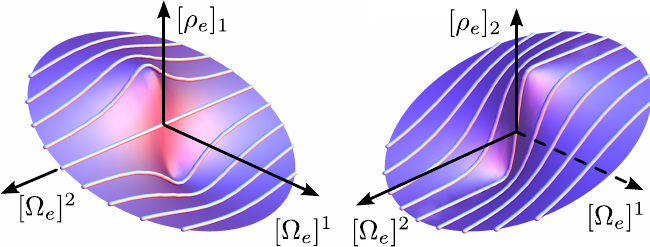}
	\caption{a family of steady-state solutions for $\rhovec$.  The (white) constant-$[\Omega_e]^1$ slices of the surface can be interpreted as reparameterized spatial solutions in which $[\Omega_e]^2$ represents the spatial- and magnetic field-dependence.}
	\label{fig:tubes}
\end{figure}

If we let $\Delta$ be spatially invariant, $\overline{\Delta}$ is unity and $[\Omega_e]^1 = [\Omega_e^0]^1$ is also spatially invariant.  The implications of this are interesting, particularly when an attempt is made to map this $\Omegavec$-solution to a $\rhovec$-solution.  It can be shown from \eqref{eq:rhoomega} and \eqref{eq:OmegaSS} that the spatial-dependence of the components of $\rhovec$ will be contained in $[\Omega_e]^2(0,r)$.  In fact, a search for a dimensionless parameter to represent the spatial- and magnetic field-dependence quickly yields that $[\Omega_e]^2(0,r)$ itself is an excellent choice.  This means that a family of spatial steady-state solutions are to be found as constant-$[\Omega_e]^1$ slices of the surfaces of $\rhovec(\Omegavec)$, as shown in \autoref{fig:tubes}, and that $[\Omega_e]^2$ can be considered to be the dimensionless spatial variable.  

For small $[\Omega_e]^1$, \eqref{eq:rhoomega} yields the expression
\begin{align}
	[\rho_e]_2(t,r) = -\tanh[\Omega_e]^2(t,r). \label{eq:rho2SmallOmega1eps}
\end{align}
This corresponds to the $[\Omega_e]^1=0$ slice of \autoref{fig:tubes}.  In the standard basis (with functional dependencies suppressed), \eqref{eq:rho2SmallOmega1eps} becomes
\begin{align}
	[\rho_\varepsilon]_2 = \mu\Delta\tanh\left(\mu\Delta\left(B[\Omega_E]^1 - [\Omega_E]^2\right)\right). \label{eq:rho2SmallOmega1e}
\end{align} 
This is a spin-temperature analog of the \emph{Langevin paramagnetic equation} of statistical mechanics, which expresses the steady-state magnetization distribution as a function of external magnetic field and temperature.  If \eqref{eq:rho2SmallOmega1e} is linearized about small $[\Omega_E]^1$ (high spin-temperature of total energy), a spin-temperature analog of \emph{Currie's law} results.

As will be discussed in \autoref{sec:fenske}, \eqref{eq:rho2SmallOmega1eps} and \eqref{eq:rho2SmallOmega1e} are also closely related to the separation literature.

\subsubsection{Low dipole-energy magnetization transport equation}\label{sec:LDEMTE}
In many applications it will be reasonable to assume small dipole-energy.  A semi-linear magnetization transport equation can be derived from the continuity equation of \autoref{prop:magnetizationtransport} by linearizing about $[\rho_e]_1 = 0$.  This model will provide insight into the conditions for separative magnetization transport (SMT).  The continuity equation and current can be linearized in $e$-basis components, with $\Delta$ constant, as
\begin{subequations}\label{eq:LDEMTE}
\begin{alignat}{3}
	\partial_t \rhovec = &\hspace{.8ex}\dstar\jvec \label{eq:seminonlinear1}\\
	\jvec = 
	&\left(
		-\Gammaoz \partial_r [\rho_e]_1
	\right)\ 
	&&e^1\otimes dr \\
	+ 
	&\left( 
		-\Gammaoz \partial_r [\rho_e]_2
		+\Gammaoz\,\Delta\,\eta
	\right)\ 
	&&e^2\otimes dr \nonumber
\end{alignat}
where
\begin{align}\label{eq:SMTfeps}
	\eta =  
	\beta
	\left(1-\left([\rho_e]_2\right)^2\right) 
\end{align}
is a dimensionless factor and 
\begin{align}
	\beta = \frac{\dB{}}{\Delta B_d}[\Omega_e]^1
\end{align}
is the dimensionless single-spin-species \emph{SMT parameter}, written in the $e$-basis.  It is more physically intuitive in $E$-basis components:
\begin{align}
	\beta = \mu\dB{}[\Omega_E]^1.\label{eq:SMTce}
\end{align}
\end{subequations}
In \autoref{sec:fenske}, we will see that this parameter is functionally related to the relative volatility parameter $\alpha$ of the mass separation literature.  After evaluating the codifferential, \eqref{eq:LDEMTE} becomes
\begin{alignat}{3}\label{eq:LDEMTE2}
	\partial_t \rhovec 
	= 
	&\left(
		\Gammaoz 
		\partial_r^2 \rhoN{1}
		+ 
		j_2 \dB{}/B_d
	\right) 
	&&e^1 \\
	+ 
	&\left( 
		\Gammaoz \partial_r^2 \rhoN{2}
		-\Gammaoz\,\Delta\, \partial_r\eta
	\right) &&e^2. \nonumber
\end{alignat}

From \eqref{eq:LDEMTE} and \eqref{eq:LDEMTE2} the necessary and sufficient conditions for SMT are derived in \autoref{sec:fenske}.
 
\subsubsection{High spin-temperature\\magnetization transport equation}\label{sec:linear}
In the literature, high spin-temperature is often assumed~\cite{Genack1975,Eberhardt2007,Ramanathan2008}.  In many applications, especially when there is little separation, this is sufficient.  High spin-temperature approximations are low-$\Omegavec$ and low-$\rhovec$ approximations, and can be derived from the model of magnetization transport (\autoref{prop:magnetizationtransport}) by a first-order power-series expansion of the continuity equation about $\rhovec = \bm{0}$.

Proceeding with this approach in the $e$-basis, the following component expression is derived:
\begin{subequations}\label{eq:GR}
\begin{align}\label{eq:continuityGR}
	\partial_t \rhovec = \dstar\jvec &
\end{align}
\vspace{-1.5\baselineskip}
\begin{alignat}{3}
	\jvec =\
	&\Gammaoz\left(
		- \partial_r [\rho_e]_1
		+ [\rho_e]_1 \frac{\Delta'}{\Delta}
	\right)\, 
	&&e^1\otimes dr \\
	+\ 
	&\Gammaoz\left( 
		- \partial_r [\rho_e]_2
		+ [\rho_e]_2 \frac{\Delta'}{\Delta}
		- [\rho_e]_1 \frac{B'}{B_d}
	\right)\, 
	&&e^2\otimes dr. \nonumber
\end{alignat}
\end{subequations}
\autoref{eq:codifferentialj} can be used to compute a component form of the differential equation \eqref{eq:continuityGR}.  If $\Delta(r)$ is assumed to be spatially homogeneous, under a change of basis, the transport equations of Genack and Redfield are recovered.\footnote{See~\cite{Genack1975}, p.~83.}  The equivalence will be discussed in \autoref{sec:GREM}.

Genack and Redfield state that their expression must be altered to account for a nonuniform density of spins per unit volume $\Delta$.  \autoref{eq:GR} is such an alteration.

\subsubsection{Comparison of the models}


\begin{figure}[t]
	\centering
	\includegraphics[width=1\linewidth]{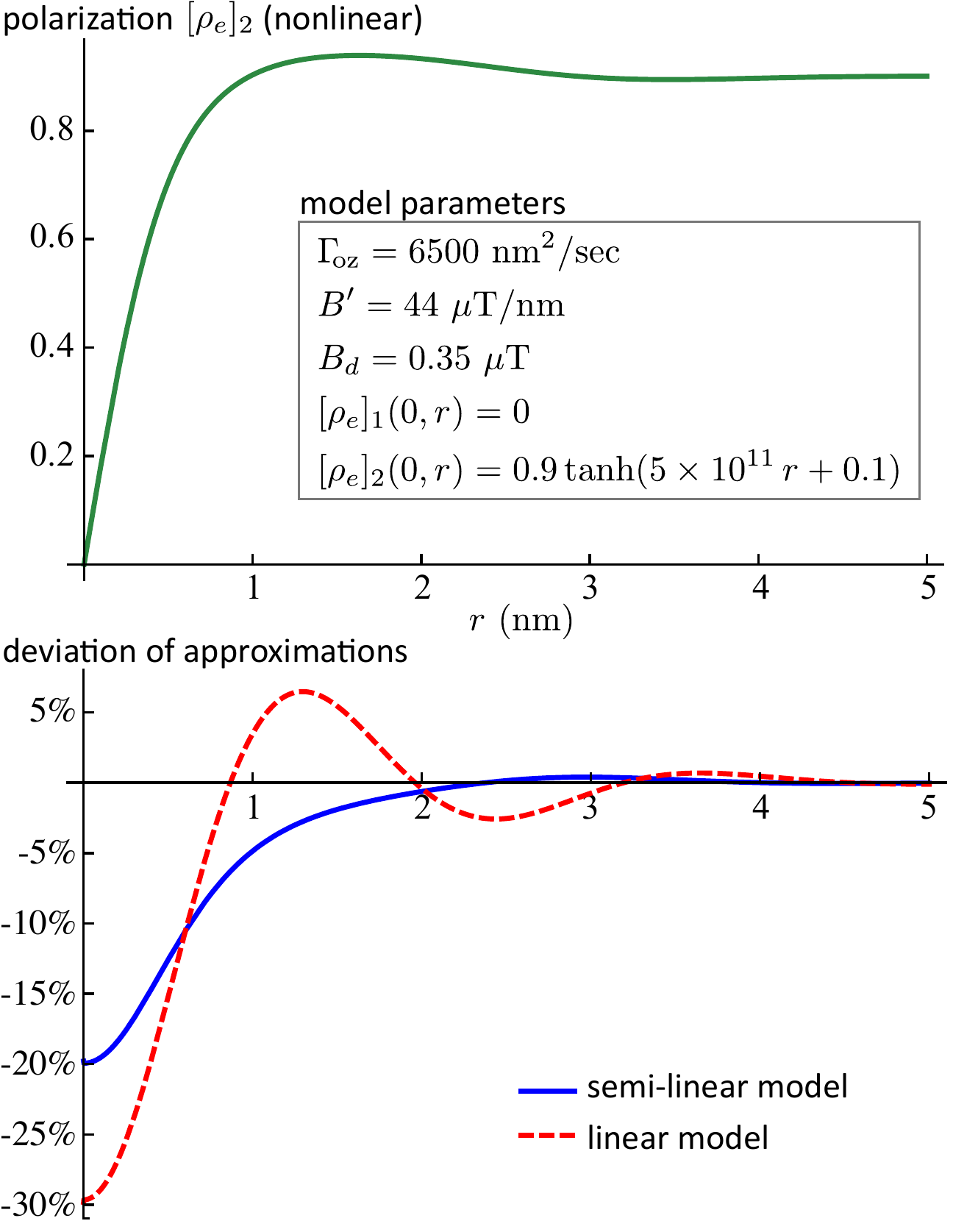}
	\caption{(top) a numerical solution for the polarization after a period of time-evolution given an initial step-like distribution to high polarization and model parameters as shown. The plot is 180 degrees rotationally symmetric about the origin. (bottom) The percent difference from our model to the semi-linear and linear approximations showing large deviations near the transition.}
	\label{fig:num}
\end{figure}

In certain regimes, the nonlinear equation of \autoref{prop:magnetizationtransport} differs from both the semi-linear \eqref{eq:seminonlinear1} and linear \eqref{eq:GR} versions.  For large $|\rhovec|$, only the nonlinear equation is valid.  For large $[\rho_e]_2$ but small $[\rho_e]_1$, only the nonlinear and partially linear equations are valid.  All three are valid for small $|\rhovec|$.

\autoref{fig:num} numerically compares the three models.  The parameters chosen are typical for a magnetic resonance force microscopy (MRFM) experiment~\cite{Degen2009,Kuehn2008,Sidles2003,Dougherty2000}.  The operating regime was large $|\rhovec|$ and field gradient, along with initial conditions of constant $[\rho_e]_1$ and step-like $[\rho_e]_2$. This models an MRFM experiment in which a method such as dynamic nuclear polarization (DNP) has hyperpolarized a sample and a region of polarization has been inverted, yielding a sharp transition in the polarization from negative to positive.  The nonlinear model predicts that, after some time, the spatial distribution of $\rhovec$ shown in the upper figure would occur.  The semi-linear and linear approximations are inaccurate in this operating regime, with the semi-linear performing better than the linear approximation, as the plot of deviations from the nonlinear model shows. The linear approximation deviates by nearly 30\% at the transition, where the semi-linear approximation deviates by approximately 20\%.

The deviations are most significant near sharp transitions in polarization. In many magnetic resonance applications this is an important regime that occurs when the polarization of a region of the sample is inverted or saturated.

\subsection{Equivalence to other models at high spin-temperatures}\label{sec:GREM}
In \autoref{sec:linear}, we claimed that if $\Delta(r)$ is assumed to be spatially homogeneous, the linear transport equation developed there is equivalent to the transport equations of Genack and Redfield.\footnote{Equations (24a,b) of~\cite{Genack1975} are the equivalent expression.  We believe that (24b) has a typo in the term containing the current (it is missing a negative sign), but it is otherwise equivalent.}  Here we show the basis transformation that yields this equivalency.  Additionally, another commonly encountered form of the linear model---one expressed in inverse spin-temperature variables---is shown to be equivalent.

Two methods have been used to verify the equivalency of the equations.  The first is to relate their basis to the $e$-basis, derive the nonlinear equation in terms of their components via \autoref{prop:magnetizationtransport}, and linearize it for small $\rhovec$-quantities.  The component transformation from the (inhomogeneous) Genack and Redfield basis to the (inhomogeneous) $e$-basis is given by the $(1,1)$-tensor transformation \mbox{$R:\mathcal{O}^*\rightarrow\mathcal{O}^*$}, which has the matrix representation
\begin{align}\label{eq:R}
	R = 
	\frac{1}{\mu\Delta}
	\begin{bmatrix}
		-B_d/\mu_0	& 0 \\
		0			& 1
	\end{bmatrix}
\end{align}
where $\mu_0$ is the magnetic constant.\footnote{See~\cite{nist}, p.~1.  The appearance of the magnetic constant is due to Genack and Redfield's definition of dipole energy-density as the magnetization times $-B/\mu_0$~\cite[p.~83]{Genack1975}.}

The second method is to directly transform the Genack and Redfield equations into \autoref{eq:GR} (with homogeneous $\Delta$) via \eqref{eq:R}.  Let $[\rho_\mathrm{gr}]_1$ and $[\rho_\mathrm{gr}]_2$ denote the magnetic susceptibility and magnetization, respectively.\footnote{We use the definitions of Genack and Redfield.}  Both the variables and the equations must be transformed by the relations
\begin{subequations}
\begin{align}
	[\rho_e]_i &= \tensor{[R]}{_i^j}[\rho_\mathrm{gr}]_j\ \text{and} \\
	\partial_t[\rho_e]_i &= \tensor{[R]}{_i^j} \partial_t[\rho_\mathrm{gr}]_j.
\end{align}
\end{subequations}
Both methods have been used to verify the equivalency.  The latter method is valid only because both bases shared the same zero-state, about which each system was linearized.

Eberhardt \emph{et al}.\ and others express this high spin-temperature model in terms of inverse spin-temperatures~\cite{Genack1975,Eberhardt2007,Ramanathan2008}.  It is tedious but straightforward to show that these are equivalent to linearized equations in $\rhovec$ that have been transformed to equations in $\Omegavec$ via a linearized version of \eqref{eq:OmegaOfRho}, under the following choice of basis: $[\rho_{\mathrm{st}}]_1$ is the dipole-energy density and $[\rho_{\mathrm{st}}]_2$ is the Zeeman energy density.  The component transformation from the standard basis to this (inhomogeneous) basis is given by the $(1,1)$-tensor transformation \mbox{$M:\mathcal{O}^*\rightarrow\mathcal{O}^*$}, which has the matrix representation
\begin{align}\label{eq:M}
	M = 
	\begin{bmatrix}
		1		& B(r) \\
		0		& -B(r)
	\end{bmatrix}.
\end{align}

The linearized map of the components of this basis to the components of its thermodynamic dual basis is given by the $(2,0)$-tensor transformation $N:\mathcal{O}^*\rightarrow\mathcal{O}$, which has the matrix representation
\begin{align}\label{eq:N}
	N = 
	\frac{-1}{\mu^2\Delta^2}
	\begin{bmatrix}
		1/B_d^2		& 0 \\
		0		& 1/B(r)^2
	\end{bmatrix}.
\end{align}
The $\rhovec$-variable equation is transformed to the $\Omegavec$-variable equation by the relations
\begin{subequations}
\begin{align}
	[\Omega_{\mathrm{st}}]^i &= \tensor{[N]}{^i^j}[\rho_\mathrm{st}]_j\ \text{and} \\
	\partial_t[\Omega_\mathrm{st}]^i &= \tensor{[N]}{^i^j} \partial_t[\rho_\mathrm{st}]_j.
\end{align}
\end{subequations}

The $\rhovec$-variable and $\Omegavec$-variable expressions of Genack and Redfield's model are both equivalent to the high spin-temperature limit of the model of magnetization transport here presented.

\section{Separative transport and the Fenske equation}\label{sec:fenske}

From \eqref{eq:LDEMTE} and \eqref{eq:LDEMTE2} it can be determined that a necessary and sufficient condition for SMT at some time $t$ and location $r$ is that $\partial_r\eta(t,r)$ is nonzero.  This condition implies several other necessary conditions.  First, it implies that $B(r)$ is necessarily spatially inhomogeneous.  Second, it implies that $|[\rho_e]_2(t,r)|$ is necessarily either less than unity or spatially varying.  Third, it implies that $[\Omega_e]^1$ is necessarily spatially inhomogeneous.  Finally, it implies that $\beta$ is necessarily spatially inhomogeneous.  If the polarization is non-unity or spatially varying, as is typically the case, the nonuniformity of $\beta$ is a both necessary and sufficient condition for SMT.  If any of these conditions is not met in a region, no SMT occurs there.


Fenske developed a set of equations to describe the process of mass separation in the fractional distillation of hydrocarbons~\cite{Fenske1932}.  These have become the standard for simple models of mass separation.\footnote{See~\cite{Kister1992}, p.~114.}  In this section, a model will be derived in the manner of Fenske that is equivalent, in some operating regimes, to the model of magnetization transport presented above.  This will highlight the separative aspect of magnetization transport.

\begin{figure}[tb]
	\centering
	\includegraphics[width=.7\linewidth]{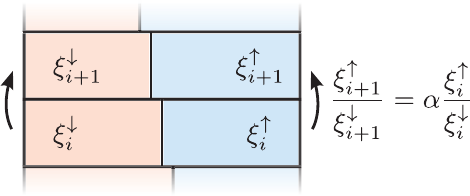}
	\caption{two ``trays'' containing fractions of up-spin $\xi^\uparrow$ and down-spin $\xi^\downarrow$.  The fraction ratio in tray $i+1$ is related to that of tray $i$ by the relative volatility $\alpha$.}
	\label{fig:column}
\end{figure}

Consider the discrete system illustrated in \autoref{fig:column}.  It is analogous to a fractional distillation column in which ``tray'' $i$ contains a certain ratio of one substance to another, and tray $i+1$ contains a greater concentration.  \autoref{fig:column} shows two ``spin'' trays in which the two ``substances'' are spin-up and spin-down, described by the fractions $\xi^\uparrow_i$ and $\xi^\downarrow_i$.  Let the total amount of spin in a tray be fixed:
\begin{align} \label{eq:xifrac}
	\xi^\uparrow_i + \xi^\downarrow_i = 1.
\end{align}

Through an exchange process among the trays, tray $i+1$ obtains a higher ratio between up- and down-spin.  In a fractional distillation column, concentration occurs by boiling liquid in a tray, the vapor of which condenses in the tray above and contains a higher concentration of the product, and fluid flows downward for mass balance.  In the spin system, the exchange occurs through dipole-dipole interactions.  Let $\alpha\ge 1$ be the \emph{relative volatility} (typically $\alpha$ is not much larger than unity).  The Fenske model describes this process by the relation\footnote{See~\cite{Halvorsen2000}, p.~35.}
\begin{align}\label{eq:fenske1}
	\frac{\xi^\uparrow_{i+1}}{\xi^\downarrow_{i+1}} = 
	\alpha\frac{\xi^\uparrow_i}{\xi^\downarrow_i}.
\end{align}

Let $\xi^\uparrow_0$ and $\xi^\uparrow_0$ be the spin fractions of some reference tray (0).  \autoref{eq:fenske1} is a difference equation that can be solved in the steady-state for the \emph{Fenske equation}\footnote{This is not the usual form of the Fenske equation, but it is equivalent.}\textsuperscript{,}\footnote{The superscript of $\alpha$ is not an index, but an exponent.}
\begin{align}\label{eq:fenske2}
	\frac{\xi^\uparrow_{i}}{\xi^\downarrow_{i}} = 
	\alpha^i\frac{\xi^\uparrow_0}{\xi^\downarrow_0}.
\end{align}
A continuum solution is found if tray $i$ is taken to be spatially small and mapped to the dimensionless spatial coordinate $\overline{r} = r \cdot \partial_r B/ B_d$ (i.e. $i\rightarrow \overline{r}$).  Polarization can be identified as
\begin{align}
	[\rho_e]_2(\overline{r}) = 2 \xi^\uparrow_{\overline{r}} - 1 = 1 - 2 \xi^\downarrow_{\overline{r}}
\end{align}
\autoref{eq:fenske2} can be written in terms of polarization.  What is more interesting, however, is that for $\alpha$ near unity, the following equation is approximate (and exact in the limit):
\begin{subequations}\label{eq:fenske3}
\begin{align}
	[\rho_e]_2(\overline{r}) = \tanh\left(a(\overline{r}-\overline{r}_0)\right)
\end{align}
where
\begin{align}
	a = \frac{\alpha-1}{\alpha+1} & & \text{and} & & \overline{r}_0 = \frac{\mathrm{arctanh}\,[\rho_e^0]_2}{a}.
\end{align}
\end{subequations}
\autoref{eq:fenske3} can be written in the standard basis by the simple relation $[\rho_\varepsilon]_2(\overline{r}) = \mu\Delta(r)[\rho_e]_2(\overline{r})$.  Comparing this to \eqref{eq:rho2SmallOmega1e} with a linear field $B(r)$, the following identification can be made:
\begin{align}
	\alpha = \frac{1+\mu B'(r) [\Omega_E^0]^1}{1-\mu B'(r) [\Omega_E^0]^1}.
\end{align}
Recalling the definition of the SMT parameter $\beta$ from \eqref{eq:SMTce}, we write
\begin{align}
	\alpha = \frac{1+\beta}{1-\beta}.
\end{align}
This result connects the theory of separative magnetization transport with Fenske-style separation theory.  For separation to occur, the relative volatility $\alpha$ must be non-unity and the SMT parameter $\beta$ must be nonzero.\footnote{As discussed in \autoref{sec:LDEMTE}, $\beta$ must be spatially inhomogeneous for SMT to occur, and so it must be nonzero over the region.}  These are two statements of the same rule.  

This highlights the separative nature of transport in the spin system.  Small concentrations of magnetization develop in the steady-state, and this can be considered separation of up-spin and down-spin.  

Beyond increasing $|B'|$, no mechanism to enhance this natural separation is apparent in this single-spin species system, as it is in a fractional distillation system, but as we will discuss in the conclusions, we believe that an enhancement of this separative effect is possible with the introduction of a second spin-species.  This will require a model that includes large-magnetization regimes and three conserved quantities (e.g. total magnetic energy, nuclear-spin magnetic moment, and electron-spin magnetic moment).  The model of \autoref{sec:magnetizationtransport} satisfies the former requirement, but it is only valid for two quantities.  However, because the framework of \autoref{sec:entropictheory} allows any number of conserved quantities, future work will develop from it a three-quantity model of magnetization transport.

\section{Conclusions and prospects}
We presented a framework for modeling the transport of any number of globally conserved quantities in any spatial configuration. We applied it to obtain a model of magnetization transport for spin-systems that is valid in new regimes (including high-polarization). Finally, we analyzed the separative quality of the magnetization transport.

Separative magnetization transport (SMT) was explored, and found to occur in a manner analogous to separative mass transport.  The analogy suggests that exploring ways of enhancing the SMT-effect may yield a useful new technique of hyperpolarizing spins.

Much is known about separative mass transport, but much remains unknown about SMT.  A distillation column has temperature and gravitational field gradients.  Its components have different volatility, and so one is more represented in the vapor phase than the other.  Understanding how these factors affect separative mass transport has been the key to harnessing separation.  For SMT, we have shown that it is necessary that the magnetic field be spatially varying.  Greater field-gradients induce greater separative transport.  Previous results have shown that diffusive transport is suppressed in a high magnetic field gradient~\cite{Eberhardt2007,Budakian2004}.  Conversely, the present results show that separative transport is \emph{enhanced}.  Certain magnetic resonance technologies are currently better-equipped than others to take advantage of this, such as magnetic resonance force microscopy (MRFM)~\cite{Degen2009,Kuehn2008,Sidles1991}, which already operates in very high magnetic field gradients.  However, typical concentrations, even in high magnetic field gradients may not be sufficient in many applications.

We conjecture from the preceding considerations that adding a second spin-species to the system may be advantageous as follows: With two spin-species (e.g. nuclear and electron), spatial gradients in the magnetization of one will affect the transport of the other through spin-spin interactions.  Using magnetic resonance techniques to locally induce such gradients in the magnetization of one species, that of the other may be concentrated.  This is unlike dynamic nuclear polarization (DNP)~\cite{Ni2013,Abragam1978,Krummenacker2012} in that no polarization is \emph{transferred} from one species to another: each species' polarization is separately conserved.  

The magnetization transport model presented in \autoref{sec:magnetizationtransport} does not include a second spin-species, but the transport framework of \autoref{sec:entropictheory} can be used to develop such a model.  Once experimentally validated, the model may be used to harness SMT to hyperpolarize and drive magnetic resonance technology development.  This work is underway and forthcoming.

\section{Acknowledgements}
	This work was supported by the Army Research Office (ARO) MURI program under contract \#~W911NF-05-1-0403.  We thank Professor John M. Lee for many insightful discussions.




\bibliographystyle{elsarticle-num}
\bibliography{/Users/picone/Dropbox/master}

\appendix*

\section{Symbol reference}
\begin{tabularx}{\linewidth}{ l@{\hskip 2pt}X}
  sym. & definition \\
  \hline
  $*$ & Hodge star operator \cite[pp.~437-8]{Lee2012} \\
  $\alpha$ & relative volatility (Eq.~\ref{eq:fenske1}) \\
  $\beta$ & separative mag. transport coef. (Eq.~\ref{eq:SMTce}) \\
  $B$ & external spatially varying magnetic field \\
  $\overline{B}$ & dimensionless $B$ (Eq.~\ref{eq:OmegaSS}) \\
  $B'$ & spatial derivative of $B$ ($\partial_r B$) \\
  $B_d$ & average dipole magnetic field \\
  $\cov$ & covariance tensor (Def.~\ref{def:covariancetensor}) \\
  $\Delta$ & spins per unit volume (Def.~\ref{def:Sentropydensity}) \\
  $\Delta'$ & spatial derivative of $\Delta$ ($\partial_r \Delta$) \\
  $\overline{\Delta}$ & dimensionless $\Delta$ (Sec.~\ref{sec:steadystatesolutions}) \\
  $d$ & exterior derivative~\cite[pp.~362-72]{Lee2012} \\
  $\dstar$ & Hodge codifferential~\cite[pp.~438-9]{Lee2012} \\
  $d r^\alpha$ & standard cotangent bundle basis \\
  $\partial_x$ & partial derivative with respect to $x$ \\
  $\partial\hspace{-.3ex}/\hspace{-.3ex}\partial r^\alpha$ & standard tangent bundle basis \\
  $\eta$ & separative mag. transport factor (Eq.~\ref{eq:SMTfeps}) \\
  $(e^i)$ & thermodynamic covector basis (Def.~
  \ref{def:Sthermodynamicbasis})\\
  $(e_i)$ & thermodynamic vector basis (Def.~
  \ref{def:Sthermodynamicbasis}) \\
  $(\varepsilon^i)$ & thermodynamic covector basis (Def.~
  \ref{def:standardcovectorbasis})\\
  $(E_i)$ & thermodynamic vector basis (Def.~
  \ref{def:thermodynamicpotentials}) \\
  $\bm{\Phi}$ & an element of $\mathcal{O}$ (Def.~\ref{def:Sentropydensity}) \\
  $\F$ & Onsager's kinetic coefficients (Def.~\ref{def:F}) \\
  $\FOZ$ & OZ-\textit{ansatz} kinetic coefficients (Def.~\ref{def:ozansatz}) \\
  $\Gammat$ & transport rate tensor (Def.~\ref{def:covariancetensor}) \\
  $\Gammaoz$ & OZ transport coefficient (Def.~\ref{def:ozansatz}) \\
  $g$ & spatial metric (Def.~\ref{def:spatialmetric}) \\
  $\G$ & entropy Hessian (Def.~\ref{def:entropyhessian}) \\
  $\jvec$ & spatial transport current (Def.~\ref{def:current}) \\
  $\mu$ & magnetic moment of an individual spin \\
  $m$ & dimension of $\U$ (Def.~\ref{def:spatialmanifold}) \\
  $M$ & basis transformation (Eq.~\ref{eq:M})
\end{tabularx}

\begin{tabularx}{\linewidth}{ l@{\hskip 2pt}X}
  $n$ & number of conserved quantities (Def.~\ref{def:conservedquantities}) \\
  $N$ & basis transformation (Eq.~\ref{eq:N}) \\
  $\circ$ & map composition \\
  $\otimes$ & tensor product~\cite[p.~306]{Lee2012} \\
  $\Omegavec$ & local thermodynamic potential (Def.~\ref{def:thermodynamicpotentials}) \\
  $\mathcal{O}$ & the set of smooth maps from $\U\times\mathbb{R}$ to $V$  \\
  $\mathcal{O}^*$ & the set of smooth maps from $\U\times\mathbb{R}$ to $V^*$ \\
  $p$ & a point on $\U$ \\
  $P$ & $\varepsilon$- to $e$-basis transform (Def.~\ref{def:Sthermodynamicbasis}) \\
  $\qvec$ & conserved thermocovector (Def.~\ref{def:conservedquantities}) \\
  $\rhovec$ & local quantity densities (Def.~\ref{def:quantitydensities}) \\
  $(r^\alpha)$ & standard spatial coordinate (Rem.~\ref{rem:spatialcoordinates}) \\
  $R$ & basis transform (Eq.~\ref{eq:R}) \\
  $\sharp$ & sharp operator~\cite[pp.~341-3]{Lee2012} \\
  $s$ & local entropy density (Def.~\ref{def:entropy}) \\
  $\bm{\Psi}$ & an element of $\mathcal{O}$ (Def.~\ref{def:Sentropydensity}) \\
  $\TU$ & tangent bundle on $\U$ \\
  $\TpU$ & tangent space at $p\in\U$ \\
  $\TStarU$ & cotangent bundle on $\U$ \\
  $\TpStarU$ & cotangent space at $p\in\U$ \\
  $\U$ & spatial manifold (Def.~\ref{def:spatialmanifold}) \\
  $\xi^\uparrow_i$ & fraction of up-spin in tray $i$ (Eq.~\ref{eq:xifrac}) \\
  $\xi^\downarrow_i$ & fraction of down-spin in tray $i$ (Eq.~\ref{eq:xifrac})
\end{tabularx}

\end{document}